\documentclass{LMCS}

\usepackage{enumerate}
\usepackage{hyperref}

\usepackage[utf8]{inputenc} 
\usepackage{algorithmic}
\usepackage{amssymb}
\usepackage{amsmath, latexsym}
\usepackage{amsfonts}
\usepackage{xspace}
\usepackage[mathscr]{eucal}
\usepackage{caption}

\newcommand{\CDlin}{{\textsc{Constant-}\textsc{Delay}$_{lin}$}\xspace}  
\newcommand\set[1]{\ensuremath{\{#1\}}\xspace}
\newcommand\cA{\ensuremath{{\mathcal A}}\xspace}
\newcommand\cN{\ensuremath{{\mathcal N}}\xspace}
\newcommand{\T}{\ensuremath{{\mathcal T}}}
\newcommand{\F}{\ensuremath{{\mathcal F}}}
\newcommand\FO{\textup{FO}\xspace}

\def\doi{7 (2:20) 2011}
\lmcsheading%
{\doi}
{1--8}
{}
{}
{Nov.~\phantom19, 2010}
{Jun.~29, 2011}
{}   
\begin{document}

\title[First-order query evaluation on structures of bounded degree]{First-order query evaluation on structures of bounded degree}

\author[W.~Kazana]{Wojciech Kazana\rsuper a}	
\address{{\lsuper a}INRIA and ENS Cachan}	
\email{kazana@lsv.ens-cachan.fr}  
\thanks{{\lsuper a}This work has been partially funded by the European
  Research Council under the European Community's Seventh Framework
  Programme (FP7/2007-2013) / ERC grant Webdam, agreement
  226513. \url{http://webdam.inria.fr/}}	

\author[L.~Segoufin]{Luc Segoufin\rsuper b} 
\address{{\lsuper b}INRIA and ENS Cachan} 
\email{see \url{http://www-rocq.inria.fr/~segoufin}} 
\thanks{{\lsuper b}We acknowledge the financial support of the Future 
  and Emerging Technologies (FET) programme within the Seventh
  Framework Programme for Research of the European Commission, under
  the FET-Open grant agreement FOX, number FP7-ICT-233599.} 


\begin{abstract}
  We consider the enumeration problem of first-order queries over
  structures of bounded degree. Durand and Grandjean have shown that
  this problem is in \CDlin. An enumeration problem belongs to \CDlin if
  for an input of size $n$ it can be solved by (i) an $O(n)$ precomputation
  phase building an index structure, followed by (ii) a phase enumerating
  the answers with no repetition and a constant delay between two
  consecutive outputs.
  In this article we give a different proof of this result based on
  Gaifman's locality theorem for first-order logic. Moreover, the
  constants we obtain yield a total evaluation time that is triply
  exponential in the size of the input formula, matching the
  complexity of the best known evaluation algorithms.
\end{abstract}

\maketitle

\section{Introduction.}

Model checking is the problem of testing whether a given sentence is true in a
given model. It's a classical problem in many areas of computer science, in
particular in verification.  If the formula is no longer a sentence but has
free variables then we are faced with the query evaluation problem. In this
case the goal is to compute all the answers of a given query on a given
database.

As for model checking, query evaluation is a problem often requiring a time at
least exponential in the size of the query. Even worse, the evaluation often
requires a time of the form $n^{O(k)}$, where $n$ is the size of the database
and $k$ the size of the query. This is dramatic, even for small $k$, when the
database is huge.

However there are restrictions on the structures that make things easier. For
instance MSO sentences can be tested in time linear in $n$ over structures of
bounded tree-width~\cite{Courcelle90} and MSO queries can be evaluated in time
linear in $n+m$, where $m$ is the size of the output of the query (note that
$m$ could be exponential in the number of free variables of the query, and
hence in $k$)~\cite{FlumFrickGrohe02}.

In this paper we are concerned with first-order logic (FO) and structures of
bounded degree. In this case the model checking problem for FO sentences is known to be
linear in $n$~\cite{Seese96}. Moreover, the constant factor is at most triply
exponential in the size $k$ of the formula~\cite{FrickGrohe04}. This last
algorithm easily extends to query evaluation obtaining an algorithm working in
time $f(k)(n+m)$ where $f$ is a triply exponential function.

As we already mentioned, the size $m$ of the output may be exponential in the
arity of the formula and therefore may still be large. In many applications
enumerating all the answers may already consume too many of the allowed
resources. In this case it may be appropriate to first output a small subset of
the answers and then, on demand, output a subsequent small number of answers
and so on until all possible answers have been exhausted. To make this even
more attractive it is preferable to be able to minimize the time necessary to
output the first answers and, from a given set of answers, also minimize the
time necessary to output the next set of answers - this second time interval is
known as the \emph{delay}.

We say that a query can be evaluated in linear time and constant delay if
there exists an algorithm consisting of a preprocessing phase taking time
linear in $n$ which is then followed by an output phase printing the answers
one by one, with no repetition and with a constant delay between each output.
Notice that if a linear time and constant delay algorithm exists then the time
needed for the total query evaluation problem is bounded by $f(k)(n+m)$ for
some function $f$. Hence this is indeed a restriction of the linear time query
evaluation algorithms mentioned above. From the best of our knowledge it is not
yet known whether a bound $f(k)(n+m)$ for some function $f$ on a query evaluation
problem implies the existence of a linear time and constant delay enumeration
algorithm. We conjecture this is not the case.

It was shown in~\cite{DurandGrandjean07} that linear time constant delay query
evaluation algorithms could be obtained for FO queries over structures
of bounded degree, hence improving the results of~\cite{Seese96}
and~\cite{FrickGrohe04}.

The proof of~\cite{DurandGrandjean07} is based on an intricate quantifier
elimination method. In this paper we provide a different proof of this result
based on Gaifman Locality of FO queries. Our algorithm can be seen as an extension
of the algorithm of~\cite{FrickGrohe04} to queries. However the index structure
built during the preprocessing phase is more complicated than the one
of~\cite{FrickGrohe04} in order to obtain the constant delay enumeration.
Moreover, our constant factor is triply exponential in the size of the formula,
while it is not clear whether the constant factor obtained
in~\cite{DurandGrandjean07} is elementary. Note that the triply exponential
constant factor cannot be significantly improved: it is shown
in~\cite{FrickGrohe04} that a constant factor only doubly exponential in the
size of the formula is not possible unless the parametrized complexity class
AW$[*]$ collapses to the parametrized class FPT.

\section{Definitions.}
\subsection{Gaifman locality and first-order logic.}

A relational signature is a tuple $\sigma=(R_{1}, \ldots, R_{l})$, each $R_i$
being a relation symbol of arity $r_i$. A relational structure over $\sigma$ is
a tuple $\cA = \left(A, R^{\cA}_{1}, \ldots, R^{\cA}_{l} \right)$, where $A =
\set{a_{1}, \ldots, a_{m}}$ is the set of elements of \cA and $R^{\cA}_{i}$ is
a subset of $A^{r_{i}}$.
We fix a reasonable encoding of structures by words over some finite
alphabet. The \emph{size} of \cA is denoted by $||\cA||$ and is the length of
the encoding of \cA.

The \emph{Gaifman graph} of a relational structure $\cA$, denoted by $G(\cA)$,
is defined as follows: the set of vertices of $G(\cA)$ is $A$ and there is an
edge $(a,b)$ in $G(\cA)$ iff there exists a relation $R_{i}$ and a tuple $t \in
R_{i}$ such that both $a$ and $b$ occur in $t$. Given $a,b \in A$, the
\emph{distance} between $a$ and $b$, denoted $\delta(a,b)$, is the length of a
shortest path between $a$ and $b$ in $G(\cA)$ or $\infty$ if $a$ and $b$ are
not connected. The \emph{distance} between two tuples $\bar a=(a_{1}, \ldots,
a_{k})$ and $\bar b=(b_{1}, \ldots, b_{l})$ of \cA, denoted $\delta(\bar a,
\bar b)$, is the $\text{min} \{ \delta(a_{i}, b_{j}) : 1 \leq i \leq k, 1 \leq
j \leq l \}$. For a given $r\in\mathbb{N}$ and a given tuple of elements $\bar
a$ of some structure $\cA$, we denote by $N_{r}(\bar a)$ the set of all elements in $A$ such
that their distance from $\bar a$ is less or equal to $r$.  The
\emph{$r$-neighborhood} of $\bar a$, denoted as $\cN_{r}(\bar a)$, is the
substructure of \cA induced by $N_{r}(\bar a)$ and expanded with one constant
for each element of $\bar a$. 
Given two tuples of elements $\bar a$ and $\bar
b$ we say that they have \emph{the same $r$-neighborhood type}, written
$\cN_r(\bar a) \simeq \cN_r(\bar b)$, if there is an isomorphism between
$\cN_{r}(\bar a)$ and $\cN_{r}(\bar b)$.

We consider first-order logic (\FO) built from atomic formulas of the form
$x=y$ or $R_{i}(x_{1}, \ldots, x_{r_{i}})$ for some relation $R_{i}$ and closed
under the usual Boolean connectives ($\neg,\vee,\wedge$) and existential and
universal quantifications ($\exists,\forall$).  When writing $\phi(\bar x)$ we
always mean that $\bar x$ are exactly the free variables of $\phi$. Given a
structure $\cA$ and a tuple $\bar a$ of elements of $\cA$, we write $\cA
\models \phi(\bar a)$ if the formula $\phi$ is true in $\cA$ after
replacing its free variables with $\bar a$. As usual $|\phi|$ denotes the size
of $\phi$.

We are now ready to state Gaifman locality for \FO.

\begin{thm}[Gaifman Locality Theorem \cite{Libkin04}]\label{hanf}
  For any first-order formula $\phi(\bar x)$, for every structure $\cA$ and
  tuples $\bar a$, $\bar b$, we have $\cN_r(\bar a) \simeq \cN_r(\bar b)$ implies
  $\cA\models \phi(\bar a)$ iff $\cA\models \phi(\bar b)$, where $r=2^{|\phi|}$.
\end{thm}

Given $d \in \mathbb{N}$, a structure is said to be $d$-degree-bounded, if
the degree of the Gaifman graph is bounded by $d$. The following nice
algorithmic property of $d$-degree-bounded structures can be proved using
Theorem~\ref{hanf}.

\begin{thm}[\cite{Seese96,FrickGrohe04}]\label{Seese}
  Fix $d \in \mathbb{N}$. The problem of whether a given $d$-degree-bounded
  structure $\cA$ satisfies a given first-order sentence $\phi$ is decidable in
  time $2^{2^{2^{O(|\phi|)}}}||\cA||$.
\end{thm}

\subsection{Model of computation and \CDlin class.}

We use Random Access Machines (RAM) with addition and uniform cost measure as a
model of computation. For further details on this model and its use in logic
see \cite{DurandGrandjean07}.

An enumeration problem is a binary relation. Given an enumeration problem $R$
and an input $x$, a \emph{solution for $x$} is a $y$ such that $(x,y) \in R$.
An enumeration problem $R$ induces a computational problem as follows: Given an
input $x$, output all its solutions. An enumeration problem is in the class
\CDlin if on input $x$ it can be decomposed into two steps:
\begin{itemize}
	\item a precomputation phase that is performed in time $O(|x|)$,
	\item an enumeration phase that outputs all the solutions for $x$ with
          no repetition and a constant delay between two consecutive
          outputs. The enumeration phase has full access to the output of the
          precomputation phase but can use only a constant total amount of extra memory.
\end{itemize}
In particular if $R$ is in \CDlin then the enumeration problem $R$ can be
solved in time $O(|x|+|\set{y: R(x,y)}|)$.  From the best of our knowledge it
is not known whether the converse is true or not. We conjecture that it is not.
More details about \CDlin can be found in \cite{DurandGrandjean07}.

\newcommand{\olex}{\ensuremath{<_{\text{lex}}}}
We are interested in the following enumeration problem for $\phi(\bar x) \in$
\FO and $d\in \mathbb{N}$: 
\begin{align*}
\text{Enum}_d(\phi)=\set{(x,y) ~:~ &x \text{ is a }
  d\text{-degree-bounded structure } \cA, y \text{ is a tuple } \bar a \text{
    of elements of }\cA\\ &\text{and } \cA\models \phi(\bar a)}
\end{align*}

We further denote by $\phi(\cA)$ the set $\set{\bar a ~:~ \cA\models \phi(\bar
  a)}$ and by $|\phi(\cA)|$ the cardinality of this set.
We show that $\text{Enum}_d(\phi)$ is in \CDlin.

\begin{thm}[\cite{DurandGrandjean07}]\label{main-result}
  There is an algorithm that for all $d\in\mathbb{N}$, all $\phi \in$ \FO
  and all $d$-degree-bounded structures \cA enumerates $\phi(\cA)$ with
  a precomputation phase taking time $2^{2^{2^{O(|\phi|)}}}\cdot||\cA||$ and
  a delay during the enumeration phase that is triply exponential in $|\phi|$.
  In particular, for all $d\in\mathbb{N}$ and all $\phi \in$ \FO the
  enumeration problem $\text{Enum}_d(\phi)$ is in \CDlin. Moreover, if the domain
  of \cA is linearly ordered, the algorithm enumerates $\phi(\cA)$ in increasing
  order relative to the induced lexicographical order on tuples.
\end{thm}

Hence the total query evaluation induced by the enumeration procedure of
Theorem~\ref{main-result} is in time $2^{2^{2^{O(|\phi|)}}}(||\cA||+|\phi(A)|)$
thus matching the model checking complexity of Theorem~\ref{Seese}.  Our
proof of Theorem~\ref{main-result} is based on Gaifman Locality Theorem while the
proof of~\cite{DurandGrandjean07} uses a quantifier elimination procedure (see
also~\cite{Lindell08} for a similar argument).
Note that it is not clear from the proof of~\cite{DurandGrandjean07} that their
algorithm is triply exponential in the size of the formula.

\section{\FO query evaluation.}

In this section we assume $d\in\mathbb{N}$ to be fixed and all our structures
are $d$-degree bounded.

A formula $\phi(\bar x)$ with $k$ free variables $\bar x=x_{1} \ldots x_{k}$ is said
to be \emph{connected around $x_1$} if $\phi(\bar x)$ logically implies
that $x_2,\ldots,x_k$ are in the $(rk)$-neighborhood of $x_1$ for $r=2^{|\phi|}$.

Let $\T_{rk}$ be the set of all isomorphism types of $(rk)$-neighborhoods of
single elements, i.e. the isomorphism types of structures of the form
$\cN_{rk}(a)$ for some element $a$ of some structure \cA. By
\emph{$(rk)$-neighborhood-type} of an element $a$ we mean the isomorphism type
of its $(rk)$-neighborhood. Because our structures are $d$-degree-bounded each
$(rk)$-neighborhood has at most $d^{rk}$ elements. For each $\tau \in \T_{rk}$ we
denote by $\mu_\tau(x)$ the fact that the $(rk)$-neighborhood-type of $x$ is $\tau$. For
each type in $\T_{rk}$ we fix a representative for the corresponding
$(rk)$-neighborhood and fix a linear order among its elements. This way, we can
speak of the first, second,\ldots, element of an $(rk)$-neighborhood. For
technical reasons, we actually fix a linear order for each $l$-neighborhood for
$l\leq rk$ such that (i) it is compatible with the distance from the center of
the neighborhood: the center is first, then come all the elements at distance
$1$, then all elements at distance $2$ and so on\ldots and (ii) the order of a
$(l+1)$-type is consistent with the order on the induced $l$-type.

For some sequence $F=\set{\alpha_2,\ldots,\alpha_{m}}$ of $(m-1)$ elements from
$[1, \ldots, d^{rk}]$, we write $\bar x=F(x_1)$ for the fact that, for
$j \in \left\{2, \ldots, m\right\}$, $x_j$ is the
$\alpha_j$-th element of the $(rk)$-neighborhood of $x_1$. Let $\F_{rk}^{m}$ be the
set of all possible such $F$. Let $\F_{rk} = \bigcup_{1 \leq m \leq k} \F_{rk}^{m}$.

For a given $\bar x = x_1 \ldots x_{k}$ a \emph{$r$-partition} of $\bar x$
is a set of pairs $\left\{ (C_{1}, F_{1}), \ldots, (C_{m}, F_{m}) \right\}$ such that
$\emptyset \neq C_{i} \subseteq \bar x$, $\bigcup_{1\leq i \leq m} C_{i} =
\left\{ x_1, \ldots, x_{k} \right\}$,  $C_{i} \cap C_{j} = \emptyset$ for
$i \neq j$, and $F_{i} \in \F_{rk}^{|C_{i}|}$. For a given $r$-partition $C$ of $\bar x$ and
$(C_{i}, F_{i}) \in C$ we write $\bar x^{i}$ to represent variables from
$C_{i}$, $x_1^i$ to represent the first variable from $C_{i}$, $x_2^i$ to represent second
variable and so on.

For a given $r$-partition $C = \left\{ (C_{1}, F_{1}), \ldots, (C_{m}, F_{m}) \right\}$
of $\bar x$ by $Div_{r}^{C}(\bar x)$ we mean a conjunction of formulas
saying that $N_r(\bar x^{i}) \cap N_r(\bar x^{j}) =\emptyset$ for all
$1 \leq i\neq j \leq m$ and formulas $\bigwedge_{(C_i,F_i) \in C} \bar x^i = F_i(x_1^i)$.
Note that the latter part implies that $\bar x^i$ is connected around $x_1^i$.

The following is an immediate consequence of
Theorem~\ref{hanf}.

\begin{lem}\label{lemma-decompose}
  Fix a structure \cA.  Then any formula $\phi(\bar x)$ with $k$ free variables
  is equivalent over \cA to a formula of the form
  \begin{equation}\label{eq-decompose}
\bigvee_{C\in C_r(\bar x)} \left[ \text{Div}_r^C(\bar x) \wedge
\bigvee_{(\tau_1, \ldots, \tau_{|C|}) \in S_C} \bigwedge_{i\leq |C|}
\mu_{\tau_{i}}(x^{i}_1)\right]
  \end{equation}
where $r=2^{|\phi|}$, $C_r(\bar x)$ is the set of all $r$-partitions of $\bar x$, 
and $S_C \subseteq (\T_{rk})^{|C|}$ is finite.
\end{lem}

\proof
Let $\phi(\bar x)$ be a formula with $k$ free variables and $r=2^{|\phi|}$.
As in the statement of this lemma, we denote by $C_r(\bar x)$ the set of all partitions
$C = \left\{ (C_{1},F_{1}), \ldots, (C_{m},F_{m}) \right\}$ of $\bar x$ with $C_{i} = \left\{
x_{1}^{i}, \ldots, x_{|C_i|}^{i} \right\}$.

By taking all possible $r$-partitions over $\bar x$ we see that
$\phi(\bar x)$ is equivalent to:
\begin{equation*}
\bigvee_{C\in C_r(\bar x)} \left( \text{Div}_{r}^{C}(\bar x) \wedge \phi(\bar x) \right)
\end{equation*}

Let $\bar a$ be a tuple of $\cA$ such that $\cA \models \phi(\bar a)$. Thus for
exactly one $C \in C_r(\bar x)$, $\cA \models Div_{r}^{C}(\bar a) \wedge
\phi(\bar a)$.  As $Div_{r}^{C}$ induces that variables from each $C_{i}$ for
some $(C_i, F_i) \in C$ are connected, the $r$-neighborhood of each $\bar a^i$
is completely included into the $(rk)$-neighborhood of $a_1^i$. Let
$m=|C|$. For $1 \leq i \leq m$ let $\tau_i$ be the $rk$-neighborhood-type of
$a_1^i$. We now take $S_C$ as the set of all such tuples $(\tau_1, \ldots,
\tau_m)$ for all tuples $\bar a$ such that $\cA\models Div_{r}^{C}(\bar a)\land
\phi(\bar a) $. By construction we have $\phi(\bar x)$ implies
\eqref{eq-decompose}. The reverse inclusion is an immediate consequence of
Gaifman Locality Theorem: When $Div_r^{C}(\bar a)$ holds, $\cN_r(\bar a^i)$ is
induced by $\cN_{rk}(a_1^i)=\tau_i$ and $F_i$. Moreover, $\cN_r(\bar a)$ is the
disjoint union of $\cN_r(\bar a^i)$ and is therefore induced by $C$. \qed

\medskip
We are now ready to prove Theorem~\ref{main-result}.
\medskip

\proof[Proof of Theorem~\ref{main-result}] Fix a formula $\phi(\bar x)$ with
$k$ free variables. Let \cA be a structure. Let $r=2^{|\phi|}$. By
Lemma~\ref{lemma-decompose}, $\phi(\bar x)$ is equivalent over \cA to a formula
of the form given by \eqref{eq-decompose}.  We assume that \cA comes with a
linear order over its elements. If not, we use the linear order induced by the
encoding of \cA.

  Intuitively the precomputation phase determines the disjunction given by
  \eqref{eq-decompose} and precomputes the
  $(rk)$-neighborhoods of each element of $\cA$. The fact that this can be done
  in time linear in $||\cA||$ and triply exponential in $|\phi|$ will make use
  of Theorem~\ref{Seese}.

  In a first step, for each $i\leq rk$ we precompute the pairs of nodes at
  distance $i$. In other words, for each $a$ in \cA, we compute the set of
  elements $b$ such that $\delta(a,b)=i$. This can easily be done in time
  linear in $rk\cdot||\cA||$ by induction on $i$: during the base case we compute
  the Gaifman graph of \cA and then we perform the classical computation of
  the transitive closure of this graph up to depth $rk$.

  In a second step, the precomputation phase computes for each element $a$ of
  $\cA$ its $(rk)$-neighborhood: for each element $a$ of \cA, we compute its
  $(rk)$-neighborhood-type and for all $i\leq d^{rk}$ a pointer from $a$ to the $i$-th
  element of its $(rk)$-neighborhood. We use an induction on the radius of the
  neighborhood to achieve this goal within the desired time constraints.
  
  As $0$-neighborhoods all share the same isomorphism type and have just one
  pointer to their centers, the induction base is obvious. So let's assume that
  in linear time in the size of $\cA$ we have computed all $l$-neighborhoods
  for all nodes.  With one more linear pass we now compute the
  $(l+1)$-neighborhoods. Fix $a \in \cA$. From the first step, we have all
  the elements of \cA at distance $l+1$ from $a$. As we already have computed
  the $l$-neighborhood, it remains to try all possible orders among those
  elements and test isomorphism with the ordered types we have initially fixed.

  There are at most $d^{l+1}$ nodes at distance $l+1$ and $l< rk$. Hence the
  number of orders we need to test is bounded by $(d^{rk})!$. Once the order is
  fixed we try all possible $(rk)$-neighborhood-types that we have initially
  fixed (there are $|\T_{rk}|$ possibilities) and then test that the two orders
  induce an isomorphism (each test simply requires going through all tuples of
  the neighborhood). Let $s(r,k,d)$ be the maximal size of a
  $(rk)$-neighborhood. Thus this step is altogether achieved in time
  $O((d^{rk})! \cdot |\T_{rk}| \cdot s(r,k,d))$ which is triply exponential in $|\phi|$
  because $r=2^{|\phi|}$, $|\T_{rk}|=O(2^{s(r,k,d)})$ and $s(r,k,d)=O(d^{rk|\sigma|})$.

  During the third step of the precomputation we determine the
  $(rk)$-neighborhood-types that are relevant for $\phi$ over \cA. Fix a $r$-partition
  $C=\set{(C_1,F_1),\ldots,(C_m,F_m)}$ of $C_r(\bar x)$ and a sequence
  $\tau_1,\ldots,\tau_m \in \T_{rk}$. This sequence is
  \emph{relevant for $C$} if $\cA \models \exists \bar x \left[\text{Div}_r^C(\bar x) \wedge
  \bigwedge_{j} \mu_{\tau_{j}}(x^{j}_1)\right]
  \wedge \phi(\bar x)$. Notice that the tests of the form
  $\mu_{\tau_{j}}(x^{j}_1)$ have been precomputed during the second step and
  can therefore now be treated as unary symbols. Similarly the tests
  $\text{Div}_r^C(\bar x)$ can be expressed using the graph computed during the
  first phase. Altogether, the first and second phase has replaced
  $\left[\text{Div}_r^C(\bar x) \wedge \bigwedge_{j} \mu_{\tau_{j}}(x^{j}_1)\right]$
  with a formula of size linear in $k$. Hence
  we can apply Theorem~\ref{Seese} in order to test whether the sequence is
  relevant for $C$ in time linear in $||\cA||$ and triply exponential in the size of
  the formula.  We do this for all possible $C$, investigating at most
  $(|\T_{rk}|)^k=2^{2^{2^{O(|\phi|)}}}$ cases. The number of possible $C$ is
  the number of possible splits of $k$ variables into disjoint and nonempty subsets
  multiplied by $(|\F_{rk}|)^k$, which altogether is again $2^{2^{2^{O(|\phi|)}}}$.
  For each $C$ we store a list of all sequences relevant for it. We call a $r$-partition
  $C$ \emph{relevant} if that list is nonempty.

  The last step of the precomputation phase orders, for each $\tau \in
  \T_{rk}$, the elements of $\cA$ having that particular
  $(rk)$-neighborhood-type and stores a pointer from one element to the next
  one according to the linear order on the elements of \cA. To do that, we just
  need to enumerate through all the elements in $\cA$, in the order provided by
  the linear order on its elements, and, using information obtained in the
  second step, add each of them to a proper list. In order to do this we need
  to be able to sort a set of elements in linear time and this can be done in
  our RAM model as explained in~\cite{Grandjean96}.

  Altogether we have a precomputation phase of the desired properties: it works
  in time linear in $|\cA|$ and triply exponential in $|\phi|$. We now turn to
  the enumeration phase.

  Fix relevant $r$-partition $C=\set{(C_1,F_1),\ldots,(C_m,F_m)}$ in $C_r(\bar x)$. We show how to
  enumerate in lexicographical order, with no repetition, constant memory and
  constant delay, all the tuples $\bar a$ such that $\cA,\bar a \models
  \text{Div}_r^C(\bar x) \wedge \bigvee_{i} \bigwedge_{j} \mu_{\tau_{ij}}(x^{ij}_1)$.
  The result will then follow from the following simple lemma, whose proof
  consist in merging two ordered lists.

  \begin{lem}[\cite{Bagan09}]\label{lemme-disjunct}
    If there is a linear order $<$ such that $R,R'$ are in \CDlin and both
    output their answers in increasing order relative to $<$, then $R\cup R'$
    is also in \CDlin and the answers can be enumerated in increasing order
    relative to~$<$. \qed
  \end{lem}

 The proof is by induction on the number $m$ of classes in the $r$-partition
 $C$. The base case being a particular case of the inductive step, we only do
 the inductive step.

 Without loss of generality we assume that the most significant variable of
 $\bar x$ is in the first variable of $\bar x^1$, that the most significant
 variable of $\bar x \setminus \bar x^1$ is the first variable of $\bar x^2$
 and so on. We simultaneously do the following for each sequence
 $\tau_1, \ldots, \tau_m$ relevant for $C$ and use Lemma \ref{lemme-disjunct}
 to avoid duplicate answers.

 Fix $\tau_1, \ldots, \tau_m$ relevant for $C$.
 Using the precomputed pointers we can enumerate one by one all elements $a_1$
 of \cA whose $(rk)$-neighborhood-type is $\tau_1$. For each such element
 let $\bar a^1=F_1(a_1)$ and we enumerate, by induction, the solutions for
 $\psi = \text{Div}_r^{C'}(\bar x) \wedge \bigvee_{i} \bigwedge_{j} \mu_{\tau_{ij}}(x^{ij}_1)$,
 where $C'$ is $C$ with $(C_1,F_1)$ removed. For each
 solution $\bar b$ obtained by induction, we check whether $N_{rk}(a_1)$
 intersects with $N_{r}(\bar b)$ or not (recall that this information has been
 precomputed during the first phase and therefore requires only constant time).
 If it does not, we have a solution
 $\bar a^1,\bar b$ for $\phi$ because of \eqref{eq-decompose}. If it does then
 we move to the next solution to $\psi$. Notice that the size of $N_{rk}(a_1)$ is bounded
 by $d^{rk}$ hence the length of false hits is bounded by $d^{rk^2}$. As we
 consider only relevant sequences of pairs, for each $\bar a^1$ we are certain to find
 at least one matching $\bar b$ that gives us a solution $\bar a^1,\bar b$ to $\phi$.
 Altogether we get the desired constant delay for the enumeration process.

 The enumeration phase needs to process all possible $r$-partitions $C$ and all
 relevant sequences of $\T_{rk}$, i.e. a number of cases triply
 exponential in $|\phi|$. Note that each such choice yields disjoint solution
 sets and can therefore be considered sequentially. Altogether this yields a
 procedure linear in the size of the output and triply exponential in $|\phi|$. \qed

\section{Conclusion}

We have given a new proof of the linear time and constant delay
enumeration problem of first-order queries over structures of bounded
degree.  Our procedure is based on Gaifman's locality theorem for
first-order logic and our constants are triply exponential in the size
of the query, and therefore induces the known complexity of the
associated model checking problem.

\section*{Acknowledgement}
The authors wish to thank Dietrich Kuske and the anonymous referees for their
constructive comments on earlier versions of this paper.

\bibliographystyle{plain}
\bibliography{bibliography}

\end{document}